\documentstyle[12pt]{article}
\begin{document}
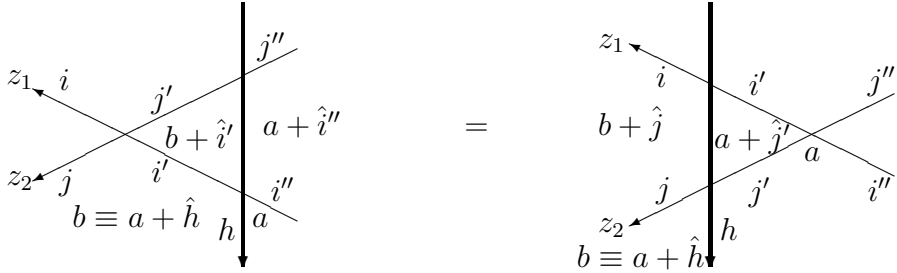
\title{\Large \bf The Dynamical Yang-Baxter Relation \\and the Minimal
Representation of the Elliptic Quantum Group}
\author{{Heng Fan$^a$, Boyu Hou$^a$, Kangjie Shi$^a$, Ruihong Yue$^a$ and Shaoyou Zhao$^b$}\\
{\small a  Institute of modern physics, Northwest university, P.O.Box 105}\\
{\small Xi'an 710069, P.R. China}\\
{\small b  Department of physics, Beijing institute
of technology}\\
{\small Beijing 100081, P.R. China}} \maketitle
\begin{abstract}
  In this paper, we give the general forms of the minimal $L$ matrix (the
elements of the $L$-matrix are $c$ numbers) associated with the
Boltzmann weights of the $A_{n-1}^1$ interaction-round-a-face
(IRF) model and the minimal representation of the $A_{n-1}$ series
elliptic quantum group given by Felder and Varchenko. The explicit
dependence of elements of $L$-matrices on spectral parameter $z$
are given. They are of five different forms (A(1-4) and B). The
algebra for the coefficients (which do not depend on $z$) are
given. The algebra of form A is proved to be trivial, while that
of form B obey Yang-Baxter equation (YBE). We also give the PBW
base and the centers for the algebra of form B.
\end{abstract}

\section {Introduction}
~~~~
 Recently, many papers  have  focused on the many-body long-distance
integrable dynamical system, such as the Ruijsenaar Schneider model and
the Calogero Moser (CM) model[1-3]. They are
closely connected with the quantum Hall effect in the condense matter
physics and the Seiberg Witten (SW) theory in the field theory, especially
for the equations of the spectral curve in the SW theory, namely, the
modified eigenvalue equations of the Lax matrices in the above integrable
models[4-6]. These Lax
matrices are the classical limit of the $L$ matrices which is associated
with the interaction-round-a-face (IRF) model of Lie
group[7-9] and the
modified Yang-Baxter relation (NSF equation)[10-13].
All these $L$-matrices are corresponding to the
representation of the elliptic quantum group which was proposed by Felder
and Varchenko\cite{fe1,fe2}.  So it is very interesting to study  the
general solution of the $L$-matrices.

In this paper, we study the simplest case of $L$-matrices which satisfy
the Dynamical Yang-Baxter Relation (DYBR) for the $A_{n-1}$ group. The
deep study of the $A_{n-1}$ group case can help us to understand the other
Lie group cases  because a subset of the other Lie groups can be
constructed by the $A_{n-1}$ group. We only study
the simplest case of $L$ matrices, that is to say, the Hilbert  space of
the $L$-operator is a scale function space. We find that the $L$ matrices
can only have five  possible forms, form A(1), A(2), A(3), A(4) and form
B. The form A(1) and B can be constructed by the factorized $L$
matrices[14-17]. And the coefficient part of form B
obeys a set of quadratic equations which can be related to the
Shibukawa-Ueno operator\cite{shibu}.
The algebra of these quadratic relations have explicit PBW base and
satisfy the YBE without spectral parameter $z$.
We find that all known
$L$-matrices\cite{jim3,fe2,has} of related
problem are equivalent to one representation of this algebra. But it is
still an open question that whether it is the unique one.

The present paper is organized as follows. In section 2, we study
the dependence of the elements of $L$-matrix with spectral
parameter $z$. In section 3, we study the dependence of the
essential part of these elements, which are functions of both $z$
and $(a,b)$, with respect to the indices of the elements. We prove
that there are five possible classes of the minimal $L$-matrices.
We then relate the elements of adjacent lattice points $(a,b)$ and
$(a+\hat i', b+\hat i)$ in the end of this section. Section 4 is
written for the equations of the coefficient part of the elements
of $L$-matrices, as a necessary and sufficient condition of
$L$-matrices to satisfy the DYBR. This leads to two kinds of
algebras. The A algebra, which is corresponding to the form A(i)
(i=1,2,3,4), is trivially commutative and thus the coefficients of
form A(i) can be determined completely. The algebra for form B
coefficients is studied further in section 5, which satisfies YBE,
and we establish the PBW base for it. We also give center elements
for this algebra. In the last section, we give a known solution to
the equations of form B. Through out this paper, we always assume
all elements of $L$-matrix are $c$ number functions which are not
identically zero and $n\ge 4$.

\section{DYBR and the relation between factorized $L$-matrix and minimal
$L$-matrix }
~~~~
It is well known that the Boltzmann weight of the  $A^{(1)}_{n-1}$
IRF\cite{jim1,jim2,jim3} model can be written as
\begin{equation}\begin{array}{ll}
 R(a|z)_{ii}^{ii}=\displaystyle\frac{\sigma(z+w)}{\sigma(w)},&
 R(a|z)_{ij}^{ij}=\displaystyle\frac{\sigma(z)\sigma(
                 a_{ij}-w)}{\sigma(w)\sigma(a_{ij})}\quad
             for\;\;i\ne j,\\
R(a|z)_{ij}^{ji}=\displaystyle\frac{\sigma(z+a_{ij})}{\sigma(a_{ij})}
           \quad  for\;\;i\ne j,&
R(a|z)_{i'j'}^{i\ j}=0 \quad\mbox{for other cases}, \label{R}
\end{array}\end{equation}
where $a\equiv(m_0,m_1,\cdots,m_{n-1})$ is an $n$-vector, and $
a_{ij}= a_i-a_j$, $a_i=w(m_i-\frac{1}{n}\sum_l m_l+w_i)$,  $m_i$
($i=0,1,\cdots,n-1$) are integers which describe the state of model, while
 $\{w,w_i\}$ are generic c-numbers which are the parameters of the model,
and $\sigma(z)\equiv\theta\left[^{\frac{1}{2}}_{\frac{1}{2}}\right](z,\tau),$
with
$$\theta\left[^a_b\right](z,\tau)\equiv
                \sum_{m\in Z}e^{i\pi(m+a)^2\tau+2i\pi(m+a)(z+b)}.$$
We define  an n-dimension vector $\hat j=(0,0,\cdots,0,1,0,\cdots)$, in
which $j$th component is 1.

We consider a matrix whose elements are linear operators. We
denote the elements of the matrix as $ L(^a_b|z)^j_i$. The $R$-matrix and
the $L$-matrix can also be depicted by the following figures,\\

\begin{picture}(50,50)(0,0)
\put(150,-20){\vector(-2,1){80}}
\put(150,20){\vector(-2,-1){80}}
\put(107,-10){$a$}\put(127,-20){$i'$}\put(87,16){$i$}
\put(127,16){$j'$}\put(87,-20){$j$}\put(65,25){$z_1$}\put(65,-26.5){$z_2$}
\end{picture}
\begin{picture}(50,50)(0,0)
\put(300,5){\vector(-1,0){80}}
\put(260,35){\vector(0,-1){60}}\put(260.5,35){\vector(0,-1){60}}
\put(261,35){\vector(0,-1){60}}
\put(230,20){$b+{\hat i}$}
\put(280,-10){$a$}\put(210,-15){$b\equiv a+{\hat h}$}\put(240,8){$i$}
\put(275,8){$j$}\put(265,-20){$h$}\put(213,2.5){$z$}
\put(270,25){$a+{\hat j}$}
\end{picture}
\\ 
\bigskip
\\
$$\begin{array}{cc}
Figure\; 1: The\;elements \;of\;R\mbox{-}matrix\;\;\;\;&\;\;\;\;
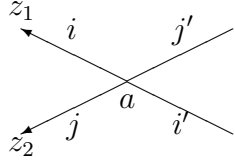
Figure\; 2: The\; element \;of\;L\mbox{-}matrix,\\
R(a|z_1-z_2)_{ij}^{i'j'}.&L(a,h|z)_i^j\equiv L(_b^a|z)_i^j.
\end{array}$$
\\
\\
$$
\begin{picture}(50,50)(0,0)
\put(-10,-32.5){\vector(-2,1){100}}
\put(-10,32.5){\vector(-2,-1){100}}
\put(-30,50){\vector(0,-1){100}}\put(-30.5,50){\vector(0,-1){100}}
\put(-31,50){\vector(0,-1){100}}
\put(-27,-35){$a$}\put(-95,-35){$b\equiv a+{\hat h}$}\put(-100,-21){$j$}
\put(-100,17){$i$}\put(-40,-40){$h$}\put(-120,19){$z_1$}
\put(-60,-5){$b+\hat i'$}
\put(-120,-19){$z_2$}\put(-20,-25){$i''$}\put(-25,30){$j''$}
\put(-23,0){$a+{\hat i''}$}\put(-65,10){$j'$}\put(-65,-18){$i'$}
\end{picture}
=
\begin{picture}(50,50)(0,0)
\put(150,-15.5){\vector(-2,1){100}}
\put(150,15.5){\vector(-2,-1){100}}
\put(80,50){\vector(0,-1){100}}\put(80.5,50){\vector(0,-1){100}}
\put(81,50){\vector(0,-1){100}}
\put(116,-10){$a$}\put(30,-50){$b\equiv a+{\hat h}$}\put(60,18){$i$}
\put(38,0){$b+\hat j$}\put(82,-5){$a+\hat j'$}
\put(60,-24){$j$}\put(84,-40){$h$}\put(38,33){$z_1$}
\put(38,-37){$z_2$}\put(140,-25){$i''$}\put(140,17){$j''$}
\put(95,15){$i'$}\put(95,-25){$j'$}
\end{picture}$$
\\
\\
\\$$ Figure\; 3: The\;\; \;dynamical \;\;Yang\mbox{-}Baxter \;\;relation.$$

The dynamical Yang-Baxter relation (DYBR) is written as (also see figure
3)
\begin{equation}
\sum_{i',j'}R(b|z_1-z_2)_{ij}^{i'j'}L(_b^a|z_1)_{i'}^{i''}L(_{b+\hat
               i'}^{a+\hat i''}|z_2)_{j'}^{j''}
 =\sum_{i',j'}L(_b^a|z_2)_{j}^{j'}L(_{b+\hat j}^{a+\hat j'}|z_1)_{i}^{i'}
                      R(a|z_1-z_2)_{i'j'}^{i''j''},\label{DYBR}
\end{equation}
where $b\equiv(m_0^b,m_1^b,\cdots,m_{n-1}^b),$
$a\equiv(m_0^a,m_1^a,\cdots,m_{n-1}^a),$ We note that
Eq.(\ref{DYBR}) gives the quadratic relation of the elements of
$L$. If we let $b=a+h$, the form of the equation will be the same
as that given in the Ref.\cite{fe1,fe2}, and the relation which
$L$ satisfies is the definition relation of the elliptic quantum
group proposed by Felder and Varchenko. Here, the elements of the
$L$-matrix are operators, and Eq.(\ref{DYBR}) is the algebra of
these operators. In this paper,  we only discuss the minimal form
of the operators, namely, we only consider the simplest case that
all elements are  c numbers. In this situation, the
$L(_b^a|z)_i^j$ is scalar functions of $(a,b,z,i,j)$. We will try
to find the general form of such $L$-matrix. From Eq.(\ref{DYBR}),
we have
\begin{eqnarray}
\frac{L(_{b+\hat i}^{a+\hat
  i'}|z_1)_{i}^{i'}}{L(_b^a|z_1)_{i}^{i'}}&=&
        \frac{L(_{b+\hat
  i}^{a+\hat i'}|z_2)_{i}^{i'}}{L(_b^a|z_2)_{i}^{i'}},\label{DYBR-1}\\
R(b|z_1-z_2)_{ii}^{ii}&=&R(a|z_1-z_2)_{i'j'}^{i'j'}
   \frac{L(_b^a|z_2)_i^{j'}}{L(_{b+\hat i}^{a+\hat i'}|z_2)_i^{j'}}
\frac{L(_{b+\hat i}^{a+\hat
j'}|z_1)_i^{i'}}{L(_b^a|z_1)_i^{i'}}\nonumber\\
&+&R(a|z_1-z_2)^{i'j'}_{j'i'}
   \frac{L(_b^a|z_2)_i^{i'}}{L(_{b+\hat i}^{a+\hat i'}|z_2)_i^{j'}}
   \frac{L(_{b+\hat i}^{a+\hat i'}|z_1)_i^{j'}}{L(_b^a|z_1)_i^{i'}}
                    \quad (i'\ne j'),\label{DYBR-2}\\
R(a|z_1-z_2)_{i'i'}^{i'i'}&=&R(b|z_1-z_2)_{ij}^{ij}
   \frac{L(_b^a|z_1)_i^{i'}}{L(_{b+\hat j}^{a+\hat i'}|z_1)_i^{i'}}
\frac{L(_{b+\hat i}^{a+\hat
i'}|z_2)_j^{i'}}{L(_b^a|z_2)_j^{i'}}\nonumber\\
&+&R(b|z_1-z_2)_{ij}^{ji}
   \frac{L(_b^a|z_1)_j^{i'}}{L(_{b+\hat j}^{a+\hat i'}|z_1)_i^{i'}}
   \frac{L(_{b+\hat j}^{a+\hat i'}|z_2)_i^{i'}}{L(_b^a|z_2)_j^{i'}}
               \quad  (i\ne j).\label{DYBR-3}
\end{eqnarray}

By solving Eq.(\ref{DYBR-2}) and Eq.(\ref{DYBR-3}), we can
determine $L(^a_b|z)_i^j$ as the function of $z$. Let
\begin{eqnarray}
&&\frac{L(_b^a|z_2)_i^{j'}}{L(_{b+\hat i}^{a+\hat
i'}|z_2)_i^{j'}}\equiv
                                   g(z_2),\;\;\;\;\;\;
\frac{L(_{b+\hat i}^{a+\hat j'}|z_1)_i^{i'}}{L(_b^a|z_1)_i^{i'}}\equiv
                                   h(z_1),\;\;\;\;\;\;
\frac{L(_{b+\hat i}^{a+\hat i'}|z)_i^{j'}}{L(_b^a|z)_i^{i'}}\equiv
                                   f(z),\\
&&\frac{R(b|z_1-z_2)_{ii}^{ii}}{R(a|z_1-z_2)_{i'j'}^{i'j'}}\equiv
                               A(z_1-z_2),\;\;\;\;\;\;\;\;\;
-\frac{R(a|z_1-z_2)_{j'i'}^{i'j'}}{R(a|z_1-z_2)_{i'j'}^{i'j'}}\equiv
                               B(z_1-z_2).
\end{eqnarray}
We then  rewrite Eq.(\ref{DYBR-2}) as
\begin{equation}
g(z_2)h(z_1)=A(z_1-z_2)+B(z_1-z_2)f(z_1)/f(z_2)\equiv F(z_1,z_2).\label{bas}
   \label{eq1-bas}
\end{equation}
We find that the left hand side  of the above equation is
factorized by the functions of $z_1$ and $z_2$. So taking logarithm
to the both sides of the above equation and taking the derivative with
respect to $z_1$ and $z_2$, we have
\begin{equation}
\frac{\partial^2}{\partial z_1\partial z_2}\ln F(z_1,z_2)=0.
\end{equation}
Hence the above equation gives
\begin{equation}
F(z_1,z_2)\frac{\partial^2}{\partial z_1 \partial z_2}F(z_1,z_2)
              -\frac{\partial}{\partial z_1}F(z_1,z_2)
               \frac{\partial}{\partial z_2}F(z_1,z_2)=0.
    \label{eq2-bas}
\end{equation}
By using Eq.(\ref{eq1-bas}) and Eq.(\ref{eq2-bas}),  we can get an
algebraic equation of 2nd order about $f(z_1)$
\begin{eqnarray}
& & f(z_1)^2\left [d_1f'(z_2)+d_2f(z_2)\right ]+f(z_1)\left
      [d_3f'(z_2)^2+d_4f'(z_2)f(z_2)+d_5f(z_2)^2\right ]\nonumber \\
&+&d_6f'(z_2)f(z_2)^2+d_7f(z_2)^3=0, \label{eq-2order}
\end{eqnarray}
where $d_i \;\; (\; i=1,\;2,\;\cdots, \; 7 \;)$ are known
functions of $z_1-z_2$. Define
\begin{eqnarray*}
y&=&\frac{f(z_1)}{f(z_2)},\\
\theta&\equiv& \frac{f'(z_2)}{f(z_2)}=\displaystyle
   \frac{\partial}{\partial z_2}
\ln\left\{\frac{L(_{b+\hat i}^{a+\hat i'}|z_2)_i^{j'}}
        {L(_b^a|z_2)^{i'}_i}  \right\}.
\end{eqnarray*}
Then, Eq.(\ref{eq-2order}) can be rewritten as
\begin{equation}
y^2(d_1\theta+d_2)+y(d_3\theta^2 +d_4\theta+d_5)+(d_6\theta+d_7)=0.
     \label{eq-2order1}
\end{equation}
When $z_2$ is fixed, the coefficients of Eq.(\ref{eq-2order1}) are
the functions of $z_1$. So $y$ is also a function of $z_1$. Since
Eq.(\ref{eq-2order1}) is of 2nd order, the $y$ can only have two
solutions $ y_1(z_1,z_2)$ and $y_2(z_1,z_2)$ at most. If we can
find two different $L$-matrices $L_1(^a_b|z)$ and $L_2(^a_b|z)$
which satisfy the DYBR with same $\theta$. we must have
$f(z_1)/f(z_2)=f_1(z_1)/f_1(z_2)$ or
$f(z_1)/f(z_2)=f_2(z_1)/f_2(z_2)$, where $f_1$ and $f_2$ are
obtained by the two different $L$'s.
 Then, we can obtain
$f(z_1)\sim f_1(z_1)$ or $f(z_1)\sim f_2(z_1)$, where $``\sim"$
implies that as the function of $z_1$, two sides of it can only be
different with a constant respect to $z_1$. Thus, we can conclude
that if there are two $L_i(^a_b|z)$ ($i=1,2$) which satisfy the
DYBR and are not proportion to each other, and when $z=z_2$, they
have same $\theta=\theta_1(z_2)=\theta_2(z_2)$, then every $f(z)$
related with $L(_b^a|z)$  satisfying $f'(z)/f(z)=\theta$ when
$z=z_2$, must  satisfy
\begin{equation}
f(z)=f_1(z)const. \quad \mbox{or}\quad f(z)=f_2(z)const. \;,\label{f(z)}
\end{equation}
where the constants do not depend on $z$.

 Now we consider the factorized $L$-matrix[14-17]
which has an adjustable parameter $\delta$. We will show that for the
given $z_2$ and $\theta$, there are generally two different $\delta$'s
which can give
$f'_{\delta_1}(z_2)/f_{\delta_1}(z_2)
        =f'_{\delta_2}(z_2)/f_{\delta_2}(z_2)=\theta$

 Considering the intertwiner of the $Z_n$ Belavin model and the
$A^{(1)}_{n-1}$ IRF model\cite{bax,jim4}, we have
\begin{eqnarray*}
\varphi_{a+\hat i,a}^{(j)}(z)&=&\theta\left[\begin{array}{c}
  \frac{1}{2}-\frac{j}{n}\\ \frac{1}{2}\end{array}\right
               ](z+n(a+\hat i)_i,n\tau)\equiv
\theta^{(j)}(nz_i),\\
(a+\hat i)_i&=&w(m_i+1-\frac{1}{n}\sum_l(m_l+\delta_{il})+w_i)
        =a_i+w(1-\frac{1}{n}).
\end{eqnarray*}
Define $\tilde\varphi_{a+\hat\nu,a}^{(j)}(z)$ which satisfies
$$
\sum_{j=0}^{n-1}\tilde\varphi_{a+\hat\nu,a}^{(j)}(z)
    \varphi_{a+\hat\mu,a}^{(j)}(z)=\delta_{\mu\nu}.$$
Let
\begin{equation}
\bar L_s(_b^a|z)^\nu_\mu
=\sum_{j=0}^{n-1}\tilde\varphi_{a+\hat\nu,a}^{(j)}(z)
    \varphi_{b+\hat\mu,b}^{(j)}(z+s),
\end{equation}
where $s$ is an arbitrary parameter. Then by using the
correspondence relation between face and vertex\cite{jim4}, we can
prove that the $L$-matrix above satisfies the DYBR
Eq.(\ref{DYBR}). After some derivation, we have\cite{shi}
$$
\bar L_s(_b^a|z)^\nu_\mu
=\frac{\sigma(z+\Delta+(n-1)w-\frac{n-1}{2}+\frac{s}{n}+b_\mu- a_\nu)}
      {\sigma(z+\Delta+(n-1)w-\frac{n-1}{2})}
\prod_{j(\ne \nu)}\frac{\sigma(\frac{s}{n}+b_\mu-a_j)}
     {\sigma(a_\nu-a_j)}
$$
with $\Delta=w\sum_jw_j$. Let
$\delta=\Delta+(n-1)w-(n-1)/2+s/n=\delta(s),\ \delta'=s/n$. Since
$\sigma(z+\Delta+(n-1)w-(n-1)/2)$ is irrelevant with
$a,b,\mu,\nu$, from the above formula, we can prove that
\begin{eqnarray}
L_\delta(_b^a|z)^\nu_\mu
&=&\bar L_s(_b^a|z)^\nu_\mu\sigma(z+\Delta+(n-1)w-\frac{n-1}{2})\nonumber\\
&=&\sigma(z+\delta+b_\mu-a_\nu)
\prod_{j\ne \nu}\frac{\sigma(\delta'+b_\mu-a_j)}{\sigma(a_\nu-a_j)} \label{factor}
\end{eqnarray}
also satisfy the DYBR (Eq.(\ref{DYBR})).

Considering the definition of $\theta$, we have
\begin{equation}
\theta(z)=\frac{f'_\delta(z)}{f_\delta(z)}
=\frac{\sigma'(z+\delta+b_i-a_{j'}+w)}{\sigma(z+\delta+b_i-a_{j'}+w)}
-\frac{\sigma'(z+\delta+b_i-a_{i'})}{\sigma(z+\delta+b_i-a_{i'})}.\label{theta}
\end{equation}
By using the properties of the $\theta$-function, one can show that for a given $\theta$,
there generally exist two different $\delta$'s satisfying Eq.(\ref{theta}).

From Eq.(\ref{f(z)}), we know that for the $L$-matrix which
satisfies the DYBR,
\begin{equation}
f(z)\sim f_\delta(z)        \label{f}
\end{equation}
must be held for certain $\delta$. And from  Eq.(\ref{bas}), we
know that $g(z)$ and $h(z)$ can be determined completely by $f(z)$
up to a scale. So we have
\begin{equation}
g(z)\sim g_\delta(z), \quad h(z)\sim h_\delta(z). \label{g}
\end{equation}
Here the parameter $\delta$ is the same as that in Eq.(\ref{f}). Then, from
Eq.(\ref{f}) and Eq.(\ref{g}), we have
\begin{eqnarray}
& &\frac{L(_{b+\hat i}^{a+\hat i'}|z)_i^{j'}}{L(_b^a|z)_i^{i'}}\sim
 \frac{\sigma(z+\delta+b_i-a_{j'}+w)}{\sigma(z+\delta+b_i-a_{i'})},
     \label{L/L-1}\\
& &\frac{L(_{b+\hat i}^{a+\hat j'}|z)_i^{i'}}{L(_b^a|z)_i^{i'}}\sim
 \frac{\sigma(z+\delta+b_i-a_{i'}+w)}{\sigma(z+\delta+b_i-a_{i'})},
      \label{L/L-2}\\
& &\frac{L(_b^a|z)_i^{j'}}{L(_{b+\hat i}^{a+\hat i'}|z)_i^{j'}}\sim
 \frac{\sigma(z+\delta+b_i-a_{j'})}{\sigma(z+\delta+b_i-a_{j'}+w)}.
      \label{L/L-3}
\end{eqnarray}
So from Eq.(\ref{L/L-1}) and Eq.(\ref{L/L-3}), we can obtain
\begin{equation}
\frac{L(_b^a|z)_i^{j'}}{L(_b^a|z)_i^{i'}}\sim
\frac{\sigma(z+\delta+b_i-a_{j'})}{\sigma(z+\delta+b_i-a_{i'})}.\label{L/L-4}
\end{equation}
In Eqs.(\ref{L/L-1})-(\ref{L/L-4}), all  $\delta$'s are the same.
We note here that the $\delta$ may depend on $i,i',j',a,b$, but it
does not depend on $z$, i.e. $\delta=\delta_i(abi'j')$. One sees
from Eqs.(\ref{L/L-1}), (\ref{L/L-2}) and (\ref{L/L-4})
$\delta_i(i'j')\cong\delta_i(j'i')$ (mod $\Lambda_\tau$).

Similarly, from the Eq.(\ref{DYBR-3}), we have
\begin{eqnarray}
& &\frac{L(_{b+\hat j}^{a+\hat i'}|z)_i^{i'}}{L(_b^a|z)_j^{i'}}\sim
\frac{\sigma(z+\delta+b_i-a_{i'}-w)}{\sigma(z+\delta+b_j-a_{i'})},
        \label{L/L-5}\\
& &\frac{L(_{b+\hat i}^{a+\hat i'}|z)_j^{i'}}{L(_b^a|z)_j^{i'}}\sim
\frac{\sigma(z+\delta+b_j-a_{i'}-w)}{\sigma(z+\delta+b_j-a_{i'})},
        \label{L/L-6}\\
& &\frac{L(_{b+\hat j}^{a+\hat i'}|z)_i^{i'}}{L(_b^a|z)_i^{i'}}\sim
\frac{\sigma(z+\delta+b_i-a_{i'}-w)}{\sigma(z+\delta+b_i-a_{i'})},
        \label{L/L-7}\\
& &\frac{L(_b^a|z)_j^{i'}}{L(_b^a|z)_i^{i'}}\sim
\frac{\sigma(z+\delta+b_j-a_{i'})}{\sigma(z+\delta+b_i-a_{i'})}.
        \label{L/L-8}
\end{eqnarray}
Here the dependence of the $\delta$'s are similar with the former.
We also have $\delta=\delta^{i'}(abij)$ and
$\delta^{i'}(ij)\cong\delta^{i'}(ji)$ (mod $\Lambda_\tau$).

\section{Dependence of elements of $L$-matrix with spectral parameter $z$}
~~ In this section, we study the dependence of $L(^a_b|z)^j_i$
with respect to $z$. It is found that there are only five possible
forms of $L$-matrices in the whole lattice. We prove this in the
following steps.

Step 1. Assume $i\ne i',\;j\ne j'$. From Eq.(\ref{L/L-4}) and
Eq.(\ref{L/L-8}), we have
\begin{eqnarray}
\displaystyle \frac{L(_b^a|z)_i^j}{L(_b^a|z)_{i'}^{j'}}&=&
\frac{L(_b^a|z)_i^j L(_b^a|z)_i^{j'}}
     {L(_b^a|z)_i^{j'}L(_b^a|z)_{i'}^{j'}}\sim
\frac{\sigma(z+\delta_i+b_i-a_{j})}{\sigma(z+\delta_i+b_i-a_{j'})}
\frac{\sigma(z+\delta^{j'}+b_i-a_{j'})}{\sigma(z+\delta^{j'}+b_{i'}-a_{j'})},
        \label{L/L-9}\\
\displaystyle \frac{L(_b^a|z)_i^j}{L(_b^a|z)_{i'}^{j'}}&=&
\frac{L(_b^a|z)_i^j L(_b^a|z)_{i'}^{j}}
     {L(_b^a|z)_{i'}^{j}L(_b^a|z)_{i'}^{j'}}\sim
\frac{\sigma(z+\delta^j+b_i-a_{j})}{\sigma(z+\delta^j+b_{i'}-a_{j})}
\frac{\sigma(z+\delta_{i'}+b_{i'}-a_j)}{\sigma(z+\delta_{i'}+b_{i'}-a_{j'})}
        \label{L/L-10}
\end {eqnarray}
giving
\begin{eqnarray}
&&\frac{\sigma(z+\delta_i+b_i-a_{j})}{\sigma(z+\delta_i+b_i-a_{j'})}
\frac{\sigma(z+\delta_{i'}+b_{i'}-a_{j'})}{\sigma(z+\delta_{i'}+b_{i'}-a_j)}
\frac{\sigma(z+\delta^{j'}+b_i-a_{j'})}{\sigma(z+\delta^{j'}+b_{i'}-a_{j'})}
\frac{\sigma(z+\delta^j+b_{i'}-a_{j})}{\sigma(z+\delta^j+b_i-a_{j})}
                 \nonumber \\
&&\equiv \frac{(1)\;(2)\;(3)\;(4)}{(1')(2')(3')(4')}\sim 1, \label{1}
\end{eqnarray}
where $$\begin{array}{ll}
\delta_i=\delta_i(a,b,j,j'),& \quad \delta^j=\delta^j(a,b,i,i'),\\
\delta_{i'}=\delta_{i'}(a,b,j,j'),&\quad
\delta^{j'}=\delta^{j'}(a,b,i,i').
\end{array}$$
Obviously in Eq.(\ref{1}), the zeroes of numerator must coincide
with those of denominator. From this fact and noticing that $a_j$
and $a_{j'}$, $b_i$ and $b_i'$ are generic complex numbers, we
analyze all cases and obtain
\begin{equation}
 \delta_i-\delta_{i'}\cong K(b_{i'}-b_i)
  \quad \mbox{and}\quad \delta^j-\delta^{j'}\cong K(a_j-a_{j'})\quad
   K=0,1,2,\label{K}
\end{equation}
where $\delta_i=\delta_i(j\ j')$, $\delta_{i'}=\delta_{i'}(j\
j')$, $\delta^{j}=\delta^{j}(i\ i')$, $\delta^{j'}=\delta^{j'}(i\
i')$ and $K=K(i\ i'\ j\ j')$. From Eq.(\ref{1}), we also have
\begin{eqnarray}
&& \delta_i\cong\delta^{j'}\cong\delta_{i'}\cong\delta^j,\quad
\mbox{when} \ K=0,\label{K=0}\\
&&\delta_i(j\ j')-\delta^{j'}(i\ i')\cong
b_{i'}-b_i+a_j-a_{j'},\quad \mbox{when} \ K=2.\label{K=2}
\end{eqnarray}

Step 2. Since the dimension $n\ge 4$, we may choose three
different $i_1, i_2, i_3$ and substitute $\{i_1, i_2\}$, $\{i_2,
i_3\}$, $\{i_1, i_3\}$ as $\{i, i'\}$ into Eq.(\ref{K}). This
leads to the conclusion that $K$ is independent of the indices
$i,i',j$ and $j'$.

These are the rules for the differences between $\delta_i(jj')$ and
$\delta_{i'}(jj')$ and for the differences between $\delta^j(ii')$ and
$\delta^{j'}(ii')$.

Step 3. Now let us study the differences between
$\delta_i(j_1j_2)$ and $\delta_{i}(j_3j_4)$. Consider different
indices $j_1,\ j_2,\ j_3$. We have from Eq.(\ref{L/L-4})
\begin{eqnarray*}
&&\frac{L(_b^a|z)_{i}^{j_1}}{L(_b^a|z)_{i}^{j_2}}
\frac{L(_b^a|z)_{i}^{j_2}}{L(_b^a|z)_{i}^{j_3}}
\sim\frac{\sigma(z+\delta_i(j_1j_2)+b_i-a_{j_1})}
        {\sigma(z+\delta_i(j_1j_2)+b_i-a_{j_2})}
     \frac{\sigma(z+\delta_i(j_2j_3)+b_i-a_{j_2})}
        {\sigma(z+\delta_i(j_2j_3)+b_i-a_{j_3})}.\\
&&LHS=\frac{L(_b^a|z)_{i}^{j_1}}{L(_b^a|z)_{i}^{j_3}}\sim
\frac{\sigma(z+\delta_i(j_1j_3)+b_i-a_{j_1})}
     {\sigma(z+\delta_i(j_1j_3)+b_i-a_{j_3})}.
\end{eqnarray*}
This implies
\begin{eqnarray*}
&&\frac{\sigma(z+\delta_i(j_1j_2)+b_i-a_{j_1})}
       {\sigma(z+\delta_i(j_1j_2)+b_i-a_{j_2})}
  \frac{\sigma(z+\delta_i(j_2j_3)+b_i-a_{j_2})}
       {\sigma(z+\delta_i(j_2j_3)+b_i-a_{j_3})}
  \frac{\sigma(z+\delta_i(j_1j_3)+b_i-a_{j_3})}
       {\sigma(z+\delta_i(j_1j_3)+b_i-a_{j_1})}\\
&&\equiv \frac{(1)\ (2)\ (3)}{(1')\ (2')\ (3')}\sim 1.
\end{eqnarray*}
From this equation, we obtain ,
\begin{equation}
\delta_i(j_1j_2)-k(a_{j_1}+a_{j_2})\cong
\delta_i(j_2j_3)-k(a_{j_2}+a_{j_3})\cong
\delta_i(j_1j_3)-k(a_{j_1}+a_{j_3})\quad k=0,1,\label{del-j}
\end{equation}
where $k=k(ij_1j_2j_3)$.

Step 4. Consider unequal $j_a, j_b, j_c, j_d$, and substitute
$\{j_a,j_b,j_c\},\{j_b,j_c,j_d\},\{j_a,j_c,j_d\}$ as
$\{j_1,j_2,j_3\}$ into Eq.(\ref{del-j}), we may show that $k$ is
also independent of the indices.

Therefore, from Eq.(\ref{K}) and Eq.(\ref{del-j}), we conclude
that one can always find a number $C$ independent of indices $i\
j\ j'$ such that
\begin{equation}
C\cong \delta_i(jj')-k_0(a_j+a_{j'})+Kb_i.\label{C}
\end{equation}
Similarly, we also find a number $D$ satisfying
\begin{equation}
D\cong \delta^j(ii')+k^0(b_i+b_{i'})-Ka_j,\label{D}
\end{equation}
where $D, K,\ k_0,\ k^0$ are independent of indices, and are fixed
for a given lattice point $(a,b)$.

Step 5. In the following, we discuss the cases $K=0$ and 2. For
$K=0$, one has from Eq.(\ref{K}), Eq.(\ref{K=0}), Eq.(\ref{C}) and
Eq.(\ref{D})
\begin{eqnarray*}
&&\delta_i(jj')\cong C+k_0(a_j+a_{j'})\cong \delta^j(ii')
   \cong D-k^0(b_i+b_{i'})\\
&\Rightarrow& D-C=k_0(a_j+a_{j'})+k^0(b_i+b_{i'}).
\end{eqnarray*}
Thus, the $k_0$ and $k^0$ must be zero since $C$ and $D$ are independent
of the indices. We have
\begin{equation}
\delta\cong C\cong D\cong\delta_i\cong\delta^j.\label{delj-deli}
\end{equation}
When $K=2$, From Eq.(\ref{K=2}), we can find a number $E$
satisfying
\begin{equation}
E\cong C\cong D  \quad\mbox{and}\quad k_0=k^0=1.\label{E}
\end{equation}

Step 6. We next study the relations for $C,D,K, k_0,K^0$ between
adjacent lattice point $(a,b)$ and $(a+\hat i', b+\hat i)$.
Eq.(\ref{L/L-1}) and Eq.(\ref{L/L-5}) intertwine two lattice
points. Notice that in Eqs.(\ref{L/L-1})-(\ref{L/L-4}) (or in
Eqs.(\ref{L/L-5})-(\ref{L/L-8})) the $\delta$'s are the same. By
using these equations, we can prove that $k, k_0,k^0$ are
unchanged for adjacent lattice points while
\begin{eqnarray}
&& C(a+\hat i',b+\hat i)-C(a,b)= C'-C\cong
-k_0w(1-\frac{2}{n})+Kw(1-\frac{1}{n}),\label{C'-C}\\
&& D(a+\hat i',b+\hat i)-D(a,b)= D'-D\cong
k^0w(1-\frac{2}{n})-Kw(1-\frac{1}{n}).\label{D'-D}
\end{eqnarray}
These equations imply that Eq.(\ref{E}) can not be realized in two
adjacent lattice points. Thus $K=2$ must be discarded.

According to $K,\ k_0,\ k^0$, when $(a,b)$ is given, the elements of the
$L$-matrix can take five forms.\\
(1). Form A(1). $K=1$, $ k_0=k^0=0$, from Eq.(\ref{L/L-4}),
Eq.(\ref{C}) and Eq.(\ref{D}), we have
\begin{eqnarray*}
\frac{L(^a_b|z)^j_i}{L(^a_b|z)^0_i}\sim
\frac{\sigma(z+\delta_i(0j)+b_i-a_j)}{\sigma(z+\delta_i(0j)+b_i-a_0)}\sim
\frac{\sigma(z+C-b_i+b_i-a_j)}{\sigma(z+C-b_i+b_i-a_0)}
=\frac{\sigma(z+C-a_j)}{\sigma(z+C-a_0)},
\end{eqnarray*}
and from Eq.(\ref{L/L-8}), we have
\begin{eqnarray*}
\frac{L(^a_b|z)^0_i}{L(^a_b|z)^0_0}\sim
\frac{\sigma(z+\delta^0(i0)+b_i-a_0)}{\sigma(z+\delta^0(i0)+b_0-a_0)}\sim
\frac{\sigma(z+D+a_0+b_i-a_0)}{\sigma(z+D+a_0+b_0-a_0)}
=\frac{\sigma(z+D+b_i)}{\sigma(z+D+b_0)}.
\end{eqnarray*}
Therefore, we obtain
\begin{eqnarray}
L(_b^a|z)_i^j&\sim&\frac{\sigma(z+C-a_j)}{\sigma(z+C-a_0)}
\frac{\sigma(z+D+b_i)}{\sigma(z+D+b_0)}  L(_b^a|z)_0^0\nonumber\\
&\sim&\sigma(z+C-a_j)\sigma(z+D+b_i)F(^a_b|z). \label{A1}
\end{eqnarray}

By using Eq.(\ref{L/L-9}), Eq.(\ref{C}) and Eq.(\ref{D}), we can
similarly derive other forms as follows,\\
(2). Form A(2). $K=1,\ k_0=0,\ k^0=1$, we have
\begin{eqnarray}
 L(_b^a|z)_i^j\sim
   \frac{\sigma(z+C-a_j)}{\sigma(z+D-b_i)}F(^a_b|z).\label{A2}
\end{eqnarray}
(3). Form A(3). $K=1,\ k_0=1,\ k^0=1$, we have
\begin{eqnarray}
 L(_b^a|z)_i^j\sim
   \frac{1}{\sigma(z+C+a_j)\sigma(z+D-b_i)}F(^a_b|z). \label{A3}
\end{eqnarray}
(4). Form A(4). $K=1,\ k_0=1,\ k^0=0$, we have
\begin{eqnarray}
 L(_b^a|z)_i^j\sim
   \frac{\sigma(z+D+b_i)}{\sigma(z+C+a_j)}F(^a_b|z).\label{A4}
\end{eqnarray}
(5). Form B. $K=0,\ k_0=0,\ k^0=0$, we have from Eq.(\ref{L/L-9})
and Eq.(\ref{delj-deli}), one obtains
\begin{eqnarray}
 L(_b^a|z)_i^j\sim
   \sigma(z+\delta+b_i-a_j)F(^a_b|z).\label{B}
\end{eqnarray}

The relation of $F(z)$ between adjacent lattice points $(a,b)$ and
$(a',b')$ are discussed in the appendix A.

In conclusion, there can at most five classes of $L$-matrices in the
whole lattice. Each of them is of the same form at all lattice points.

We must check that if these inductive relations are integrable in the
whole lattice. That is, if one goes from $(a,b)$ to
$(a''=a+\hat i'+\hat j',b''=b+\hat i+\hat j)$ via different paths, the
resulting $C''\ D''\ F''(z)$ should be the same. The conclusion is
affirmative.

For $a\equiv (m_0,m_1\cdots,m_{n-1}),$ define $m\equiv \sum_i
m_i$. Thus $m(a'=a+\hat i',b'=b+\hat i)=m(a,b)+1.$ We can express
five forms as follows, which satisfy all relations of adjacent lattice points,\\
(1). Form A(1). Let $C=C_0+mw(1-1/n),\ D=D_0-mw(1-1/n)$. Then
\begin{equation}
L(^a_b|z)^l_k\sim\sigma(z+C_0+mw(1-\frac{1}{n})-a_l)
                 \sigma(z+D_0-mw(1-\frac{1}{n})+b_k)F_0(z)
\end{equation}
and $C_0,\ D_0,\ F_0(z)$ are unchanged in the whole lattice.\\
(2). Form A(2). Let
\begin{eqnarray*}
C&=&C_0+mw(1-\frac{1}{n}),\quad\quad D=D_0-m\frac{w}{n},\nonumber\\
F(z)&=&F_0(z)\prod_{j=0}^{n-1}\sigma(z+D_0-m\frac{w}{n}-b_j).
\end{eqnarray*}
We then have
\begin{eqnarray*}
\frac{F'(z)}{F(z)}=
\frac{\sigma(z+D_0-(m+1)\frac{w}{n}-b_i-w(1-\frac{1}{n}))}
     {\sigma(z+D_0-m\frac{w}{n}-b_i)} =
\frac{\sigma(z+D-b_i-w)}{\sigma(z+D-b_i)}.
\end{eqnarray*}
Thus,
\begin{eqnarray}
L(^a_b|z)^l_k\sim\sigma(z+C_0+mw(1-\frac{1}{n})-a_l)\prod_{j(\ne k)}
\sigma(z+D_0-m\frac{w}{n}-b_j)F_0(z)
\end{eqnarray}
and $C_0,\ D_0,\ F_0(z)$ are unchanged in the whole lattice.\\
(3). Form A(3). Let
\begin{eqnarray*}
C&=&C_0+m\frac{w}{n},\quad\quad D=D_0-m\frac{w}{n},\\
F(z)&=&F_0(z)\prod_{j=0}^{n-1}\sigma(z+C_0+m\frac{w}{n}+a_j)
\sigma(z+D_0-m\frac{w}{n}-b_j).
\end{eqnarray*}
We then have
\begin{eqnarray*}
\frac{F'(z)}{F(z)}=
\frac{\sigma(z+C+a_{i'}+w)}{\sigma(z+C+a_{i'})}
\frac{\sigma(z+D-b_i-w)}{\sigma(z+D-b_i)}.
\end{eqnarray*}
Thus,
\begin{eqnarray}
L(^a_b|z)^l_k\sim\prod_{j(\ne l)}\sigma(z+C_0+m\frac{w}{n}+a_j)
\prod_{j(\ne k)}\sigma(z+D_0-m\frac{w}{n}-b_j)F_0(z)
\end{eqnarray}
and $C_0,\ D_0,\ F_0(z)$ are unchanged in the whole lattice.\\
(4). Form A(4). Let
\begin{eqnarray*}
C&=&C_0+m\frac{w}{n},\quad\quad D=D_0-mw(1-\frac{1}{n}),\\
F(z)&=&F_0(z)\prod_{j=0}^{n-1}\sigma(z+C_0+m\frac{w}{n}+a_j).
\end{eqnarray*}
We then have
\begin{eqnarray}
L(^a_b|z)^l_k\sim\sigma(z+D_0-mw(1-\frac{1}{n})+b_k)\prod_{j(\ne l)}
{\sigma(z+C_0+m\frac{w}{n}+a_j})F_0(z)
\end{eqnarray}
and $C_0,\ D_0,\ F_0(z)$ are unchanged in the whole lattice.\\
(5). Form B.
\begin{equation}
L(^a_b|z)^l_k\sim\sigma(z+\delta_0+b_k-a_l)F_0(z)
\end{equation}
and $\delta_0, F_0(z)$ are unchanged in the whole lattice.

Thus we can establish the $L$-matrix in the whole lattice,
if we can properly choose the coefficients of the elements of $L$-matrix.
We will discuss this problem in the next section.

\section{The coefficients irrelevant with $z$ of the elements of
$L$-matrix }
~~~~
In this section, we study the sufficient condition of $L$-matrices
for DYBR and derive the equations satisfied by  the coefficients
irrelevant with $z$ of the elements of $L$-matrix.

As an example, we study the form B which is useful in the later.
From the Eq.(\ref{B}) for the form B, The $L$-matrix takes the
form
\begin{eqnarray}
&&
L(^a_b|z)^j_i=(^a_b)^j_i\sigma(z+\delta+b_i-a_j)F(z),\\
 &&L(^{a+\hat i'}_{b+\hat i}|z)^{j'}_j=(^{a+\hat i'}_{b+\hat i})^{j'}_j
\sigma(z+\delta+b'_j-a'_{j'})F(z).
\end{eqnarray}
Then, substituting the above equation  and the Eq.(\ref{R}) for
the $R$-matrix into the DYBR Eq.(\ref{DYBR}) and noticing the fact
$$ a'_{j'}=a_{j'}+w(\delta_{i'j'}-{1\over n}),\quad
 b'_j=b_j+w(\delta_{ij}-{1\over n}),  \quad
 (\mbox{ for }a'=a+\hat i',\  b'=b+\hat i)
$$
we obtain the equations for the coefficients:
\begin{eqnarray}
& & \left(\begin{array}{l}a\\b\end{array}\right)^{i'}_i
\left(\begin{array}{l} a+\hat i' \\ b+\hat i \end{array}\right)^{i'}_i=
      \left(\begin{array}{l}a\\b\end{array}\right)^{i'}_i
\left(\begin{array}{l} a+\hat i' \\ b+\hat i
\end{array}\right)^{i'}_i, \label{B1}
\end{eqnarray}
which is trivially satisfied, and
\begin{eqnarray}
(^a_b)^{i'}_i(^{a+\hat i'}_{b+\hat i})^{j'}_i
-\frac{\sigma(a_{i'j'}-w)}{\sigma(a_{i'j'}+w)}
 (^a_b)^{j'}_i(^{a+\hat j'}_{b+\hat i})^{i'}_i=0 \quad (i'\neq j'),\label{sig-B1}
\end{eqnarray}
\begin{eqnarray}
(^a_b)^{i'}_i(^{a+\hat i'}_{b+\hat i})^{i'}_j
-(^a_b)^{i'}_j(^{a+\hat i'}_{b+\hat j})^{i'}_i=0\quad (i\neq j),
                    \label{sig-B2}
\end{eqnarray}
\begin{eqnarray}
&&(^a_b)^{i'}_j(^{a+\hat i'}_{b+\hat j})^{j'}_i
       \sigma(a_{i'j'}+b_{ij})\sigma(w)
+ (^a_b)^{i'}_i(^{a+\hat i'}_{b+\hat i})^{j'}_j
       \sigma(b_{ij}-w)\sigma(a_{i'j'})                \nonumber \\
&&-\;(^a_b)^{j'}_j(^{a+\hat j'}_{b+\hat j})^{i'}_i
       \sigma(a_{i'j'}-w)\sigma(b_{ij})=0 \quad  (i\neq j,\ i'\neq j'), \label{sig-B3}
\end{eqnarray}
respectively. In the derivation, we have used  the addition
formula
\begin{eqnarray}
&& \sigma(u+x)\sigma(u-x)\sigma(v+y)\sigma(v-y)
        -\sigma(u+y)\sigma(u-y)\sigma(v+x)\sigma(v-x)\nonumber\\
&& \;\;=\sigma(u+v)\sigma(u-v)\sigma(x+y)\sigma(x-y),\label{add}
\end{eqnarray}

Define
\begin{eqnarray}
&&(^a_b)^{i'}_i\times \prod_{l(\ne i')}\sigma(a_l-a_{i'})=[^a_b]^{i'}_i,
              \nonumber\\
&&[^a_b]^{i'}_i[^{a+\hat i'}_{b+\hat i}]^{j'}_j=Y^{i'j'}_{i\;j}.
\end{eqnarray}
Then for form B, we rewrite the Eqs.(\ref{sig-B1})-(\ref{sig-B3})
as
\begin{eqnarray}
&&Y^{i'j'}_{i\;i}-Y^{j'i'}_{i\;i}=0 \quad (i'\ne j'),
            \label{coff-B1}\\
&&Y^{i'i'}_{i\;j}-Y^{i'i'}_{j\;i}=0 \quad(i\ne j),
            \label{coff-B2}\\
&&\sigma(w)\sigma(a_{i'j'}+b_{ij})Y^{i'j'}_{j\;i}
+\sigma(a_{i'j'})\sigma(b_{ij}-w)Y^{i'j'}_{i\;j}\nonumber\\
&&-\ \sigma(a_{i'j'}+w)\sigma(b_{ij})Y^{j'i'}_{j\;i}=0 \quad (i\ne
j,\ i'\ne j').
            \label{coff-B3}
\end{eqnarray}

With same procedure, one can also show that all A forms (form
A(1)-A(4)) share a common coefficient relations
\begin{eqnarray}
&&Y^{i'j'}_{i\ i}-\frac{\sigma(a_{i'j'}-w)}
       {\sigma(a_{i'j'}+w)}Y^{j'i'}_{i\ i}=0
 \quad (i'\ne j'),
               \label{coff-A1}\\
&&Y^{i'i'}_{i\ j}-Y^{i'i'}_{j\ i}=0 \quad (i\ne j),
                \label{coff-A2}\\
&&Y^{i'j'}_{j\ i}=Y^{i'j'}_{i\ j}=\frac{\sigma(a_{i'j'}-w)}
       {\sigma(a_{i'j'}+w)}Y^{j'i'}_{j\ i} \quad (i\ne j,\  i'\ne
       j').
                \label{coff-A3}
\end{eqnarray}

For the coefficients of form A(i) (i=1,2,3,4), we can easily find
the rule. Consider a function $G(a,b)$ on a lattice points
$(a=\sum_jm^a_j\hat j,b=\sum_im^b_i\hat i)$. From the lattice
$(a,b)$, by using the relation $G(a+\hat i',b+\hat
i)=G(a,b)[^a_b]^{i'}_i,$  we can construct the function on  the
other lattice point. Because of the
Eqs.(\ref{coff-A1})-(\ref{coff-A3}), we can obtain same $G(a+\hat
i'+\hat j',b+\hat i+\hat j)$ through different paths from $(a,b)$
to $(a+\hat i'+\hat j',b+\hat i+\hat j)$. So this procedure is
integrable. This implies that there must exist a function $G(a,b)$
which can determine $[^a_b]^{i'}_i$ with
\begin{equation}
[^a_b]^{i'}_i=G(a+\hat i',b+\hat i)/G(a,b).
\end{equation}
Hence, we can solve the problem of form A completely. However, to the
coefficients of the form B, its rule is more complicated and we will
discuss it in the next section.

Obviously, if we take a gauge transformation
$$ [^a_b]^j_i\longrightarrow
\overline {[^a_b]}^j_i=[^a_b]^j_i\frac{g(a+\hat j,b+\hat
i)}{g(a,b)},$$ and if $[^a_b]^j_i$ satisfies
Eqs.(\ref{coff-A1})-(\ref{coff-A3}), $\overline {[^a_b]}^j_i$ also
satisfies these equations. In this sense, all form A coefficients
are gauge equivalent to a constant.

\section{The algebra for  form (B) coefficients}
\subsection{The PBW base of the algebra}
~~
In this section, we give the PBW base of the algebra for  form (B)
coefficients. The main result is Theorem 1. We also give the center of
this algebra. Eqs.(\ref{coff-B1})-(\ref{coff-B3}) can
be regarded as the algebraic relations
which are satisfied by the operators in the lattice
$(a=\sum_{j=0}^{n-1}m^a_j\hat j, b=\sum_{i=0}^{n-1}m^b_i\hat i)$.
We define a new operator
\begin{equation}
A^{i'}_i\equiv [^a_b]^{i'}_i\Gamma^{i'}_i,
\end{equation}
where
\begin{equation}
 \Gamma^{i'}_if(a,b)=f(a+\hat i',b+\hat i)\Gamma^{i'}_i.
\end{equation}
Namely, we regard the $a,b$ as operators, $\Gamma^{i'}_i$ is not
commutative with the function of $a,b$. In this way,  we have the
following exchange relations of the operators $\{A^{i'}_i\}$
\begin{eqnarray}
(a)& &A^{i'}_iA^{j'}_i=A^{j'}_iA^{i'}_i\quad\quad (i'\ne j'), \nonumber\\
(b)& &\sigma(a_{i'j'}+w)\sigma(b_{ij})A^{j'}_jA^{i'}_i \nonumber\\
   &=&\sigma(a_{i'j'})\sigma(b_{ij}-w)A^{i'}_iA^{j'}_j
     +\sigma(w)\sigma(a_{i'j'}+b_{ij})A^{i'}_jA^{j'}_i
      \ (i\ne i',\  j\ne j'), \label{coff-A}\\
(c)& &A^{i'}_iA^{i'}_j=A^{i'}_jA^{i'}_i \quad\quad (i\ne
j).\nonumber
\end{eqnarray}
These equations are equivalent relations to the Felder and
Varchenko's elliptic quantum algebra under  special condition. It
is worth noting that in the Eq.(\ref{coff-A}b), the coefficients
should be regarded as the functions of operators and they do not
commute with $A^j_i$. These equations are irrelevant with the
parameter $z$. This situation is similar to the relation between
the Sklyanin algebra[21-25] and the YBR of the Belavin
model[26-28]. In formulation, Eq.(\ref{coff-A}b) is also similar
to the function $R$-matrices given by Shibukawa and
Ueno\cite{shibu}.

Using the (a) and (b) of Eq.(\ref{coff-A}), we can exchange the
order of the up-indices of a pair of operators $A^{j'}_iA^{i'}_j$.
So $A^{j'}_iA^{i'}_j$ can be written as  linear combination of
$A^{i'}_iA^{j'}_j$ and $A^{i'}_jA^{j'}_i$. Therefore, we can write
the product of three operators $A^{i'}_iA^{j'}_jA^{k'}_k$ as the
linear combination of $A^{k'}_\cdot A^{j'}_\cdot A^{i'}_\cdot$.
This procedure can be done in two different ways. For the two
ways, by using Eqs.(\ref{coff-B1})-(\ref{coff-B3}), we can show
that according to the rules Eq.(\ref{coff-A}a) and
Eq.(\ref{coff-A}b) ( we will simplify it as (ab)), if the product
of three operators $A^{i'}_iA^{j'}_jA^{k'}_k$ is changed to the
linear combination of $A^{k'}_\cdot A^{j'}_\cdot A^{i'}_\cdot$ by
two different paths, their results  are equal. The paths are as
follows:
\begin{eqnarray*}
(A)&& i'j'k'\longrightarrow i'k'j'\longrightarrow k'i'j' \longrightarrow
           k'j'i',\\
(B)&& i'j'k'\longrightarrow j'i'k' \longrightarrow j'k'i' \longrightarrow
           k'j'i'.
\end{eqnarray*}
In the above transformation, we think that the result of the (ab)
transformation on two adjacent operators with same up-indices does not
change the order of them, namely,
$A^{i'}_iA^{i'}_j\Rightarrow A^{i'}_iA^{i'}_j$. And we
think the associative and the distributive law are satisfied in the
transformation.

Further more, if we consider the rule Eq.(\ref{coff-A}c), the
linear expansions of  operator products $A^{i'}_iA^{j'}_jA^{j'}_k$
and $A^{i'}_iA^{j'}_kA^{j'}_j$ by $A^{j'}_\cdot A^{j'}_\cdot
A^{i'}_\cdot$ via the (ab) transformation are  equal. Therefore,
we also call this fact Yang-Baxter equation (YBE).

Similarly,  after $A^{j'}_iA^{j'}_jA^{i'}_k$ and the
$A^{j'}_jA^{j'}_iA^{i'}_k$ change to the linear combination of the
$A^{i'}_\cdot A^{j'}_\cdot A^{j'}_\cdot $ by (ab), these two
expansion are equal via the rule Eq.(\ref{coff-A}c).

For the coefficient algebra (or Yang-Baxter algebra) which we
discussed above, we will give it a PBW base in the following. We
first give some definitions for establishing the base.

{\bf Definition 1: Bunch.} {\sl A bunch is a polynomial (or
monomial) of operator $A$'s, in which all terms has the same
number of $A$'s and the upper indices of $A$'s in all terms are
arranged in the same way.}

{\bf Example:} $$ B=\sum_{i_1i_2i_3i_4}C_{i_1i_2i_3i_4}
      A^{j_1}_{i_1}A^{j_2}_{i_2}A^{j_3}_{i_3}A^{j_4}_{i_4}$$
is a bunch. A polynomial is always a bunch.

{\bf Definition 2: Inverse order number.} {\sl To any two integers
$i',j'$ with a given order, we call the inverse order number is 1
if $i'>j'$, is 0 if $i'\le j'$. And the inverse order number of a
successive product $A^{i'}_\cdot A^{j'}_\cdot A^{k'}_\cdot\cdots$
is the sum of the inverse order numbers of all up-index pairs.}

{\bf Definition 3: Normal order product}. {\sl The (ab) normal
order product is a successive product of operators in  which the
up-indices are arranged from smaller to bigger when inspecting
from the left to the right, while the arrangement of the
down-indices can be arbitrary. The (abc) normal order product is
that the up-indices are arranged from the smaller to the bigger
and the down-indices of the operators with the same up-indices are
also arranged from smaller to bigger. Their inverse order numbers
are zero.}

{\bf Example:} $A^1_2A^1_1A^2_1A^2_3A^3_5A^4_1A^5_3A^5_1$ is an
(ab) normal order product but is not an (abc) normal order
product. By using the rule  Eq.(\ref{coff-A}c), we can change it
to the (abc) normal order product
$A^1_1A^1_2A^2_1A^2_3A^3_5A^4_1A^5_1A^5_3$.

{\bf Definition 4: Normal order expansion}. {\sl The (ab) normal
order expansion of a polynomial of $A$'s is a procedure in which
we change each term of the polynomial into a bunch of (ab) normal
order products by only using rules Eq.(\ref{coff-A}a) and
Eq.(\ref{coff-A}b). We also call the final resulting polynomial as
the  (ab) normal expansion of the original polynomial.

The (abc) normal order expansion is a procedure, in which we first
perform the (ab) normal order expansion and then we rearrange each
term of the resulting polynomial into (abc) normal order product
by using rule Eq.(\ref{coff-A}c). We also call the final result as
an (abc) normal expansion of the original polynomial.}

Then, we have a theorem.

{\bf Theorem 1:}  {\sl Transforming on a polynomial of operators
$A^j_i$ by using the rules (abc) of Eq.(\ref{coff-A}) does not
change its (abc) normal order expansion. }\\

It is worth noting that the coefficients of the expansions are
functions of the parameters $\{a,b\}$, they do not commute with
operator $A^j_i$.

The detailed proof of the theorem will be given in the appendix B.

{\bf Corollary:} {\sl The (abc) normal order products are linearly
independent.}

{\bf Proof:} If there were a linear relation $\sum C_ig_i=0$,
where $g_i$ are (abc) normal order products. The LHS must be able
to be changed to zero via Eq.(\ref{coff-A}). However, each
operation does not change the (abc) expansion. Thus it is
impossible since $C_i$ are not all zero. $\quad\quad {\bf \Delta}$

Thus the set of all (abc) normal order products is the PBW base of
the algebra defined by Eq.(\ref{coff-A}).

\subsection{The center of the algebra}
~~ By standard procedure, we may obtain the center of elliptic
quantum group (the detail will be given elsewhere).
$$ I=\frac{\Delta(a)}{\Delta(b)}\mbox{Det}\ L(^a_b|z),$$
where $\Delta(a)=\prod_{i<j}\sigma(a_{ij}),\
\Delta(b)=\prod_{i<j}\sigma(b_{ij})$,
\begin{eqnarray*}
&&\mbox{Det}\ L(^a_b|z)=\sum_P
(-1)^{\left[\mbox{Sign}P(^{0\ 1\ \cdots\
n-1}_{\mu_0\mu_1\cdots\mu_{n-1}})\right]}\\
&&\times\ L(^a_b|z)^0_{\mu_0}L(^{a+\hat 0}_{b+\hat\mu_0}|z+w)^1_{\mu_1}
\cdots L(^{a+\hat 0+\hat 1+\cdots+\hat{n-2}}
_{b+\hat\mu_0+\hat\mu_1+\cdots+\hat\mu_{n-2}}|z+(n-1)w)
^{n-1}_{\mu_{n-1}},
\end{eqnarray*}
and  $P$'s are permutations of integers $0,\ 1,\ \cdots,\ n-1$.
This agrees with that of Ref.\cite{fe2} for $n=2$.

In the case of
$$L(^a_b|z)^{i'}_{i}=\sigma(z+\delta+b_i-a_{i'})A^{i'}_i,$$
the quantum determinant can be written as
\begin{eqnarray*}
&&I(^a_b|z)=\sum_P(-1)^{\left[\mbox{Sign} P(^{0\ 1\ \cdots\
n-1}_{\mu_0\mu_1\cdots\mu_{n-1}})\right]}\\
&&\times\ \sigma(z+\delta+b_{\mu_0}-a_0)
   \sigma(z+w+\delta+b_{\mu_1}-a_1)\cdots \\
&&\times\  \sigma(z+(n-1)w+\delta+b_{\mu_{n-1}}-a_{n-1})
 A^0_{\mu_0}A^1_{\mu_1}\cdots A^{n-1}_{\mu_{n-1}}.
\end{eqnarray*}
It is easy to check
$$\Phi(z)_{\mu_0\cdots\mu_{n-1}}\equiv \sigma(z+\delta+b_{\mu_0}-a_0)
\cdots \sigma(z+(n-1)w+\delta+b_{\mu_{n-1}}-a_{n-1})    $$
is quasi doubly periodic in $\tau$ and 1:
\begin{eqnarray*}
&&\Phi(z+1)=(-1)^n\Phi(z),\\
&&\Phi(z+\tau)\\
&=&\exp\left[-2\pi i\left(\frac{n\tau}{2}+n\delta+nz+\frac{n(n-1)}{2}w
+\frac{n}{2}+\sum_ib_{\mu_i}-\sum_ia_i\right)\right]\Phi(z) \\
&=&\exp\left[-2\pi i\left(\frac{n\tau}{2}+n\delta+nz+\frac{n(n-1)}{2}w
+\frac{n}{2}\right)\right]\Phi(z)
\end{eqnarray*}
for all $\mu_0,\cdots,\mu_{n-1}$ being a permutation of
$(0,1,\cdots,n-1)$. Therefore, from a theorem of such function (see D.
Mumford, Tata Lectures on Theta, Birkhauser 1983), we have
\begin{equation}
\Phi(z)_{\mu_0,\cdots,\mu_{n-1}}=
 \sum_{i=0}^{n-1}C^i_{\mu_0,\cdots,\mu_{n-1}}f_i(z),
\end{equation}
where $\{f_i(z)\}$ are  base functions of the space of such quasi double
periodic function. For example, we may choose
$$f_i(z)=\theta\left[^{\frac{1}{2}-\frac{i}{n}}_{\ \ \frac{1}{2}}\right]
\left(nz+n\delta+\frac{n(n-1)w}{2}+\frac{n-1}{2},\ n\tau\right).$$
One can obtain $C^i_{\mu_0,\cdots,\mu_{n-1}}$ by choosing
$n$ points $z_1,\cdots,z_n$ in the above equation and solve a set of $n$
linear equations. We then derive the $n$ center elements for the algebra.
From
\begin{eqnarray*}
&&I(^a_b|z)\\
&=&\sum_i f_i(z)\left\{
\sum_P (-1)^{\left[\mbox{Sign}P(^{0\ 1\ \cdots\
n-1}_{\mu_0\mu_1\cdots\mu_{n-1}})\right]}C^i_{\mu_0,\cdots,\mu_{n-1}}
A^0_{\mu_0}A^1_{\mu_1}\cdots A^{n-1}_{\mu_{n-1}}\right\}\\
&\equiv& \sum_i f_i(z)J_i,
\end{eqnarray*}
we see that $[\Delta(a)/\Delta(b)]J_i$ are the center elements of
the algebra.

\section{A known solution for the form B coefficients}
~~
 The equations (Eqs.(\ref{coff-B1})-(\ref{coff-B3})) of form B coefficients
 seem simple but they interrelate the values of the coefficients
$[^a_b]^{i'}_i$ between
different lattice points.  To our best knowledge,  we only know the
analytic solution
\begin{equation}
[^a_b]^{i'}_i=\prod_{j(\ne i')}\sigma(\delta'+b_i-a_{j'}),
             \label{solution}
\end{equation}
which can be derived by the factorized operator of Eq.(\ref{factor})
\begin{eqnarray*}
L_\delta(^a_b|z)^{i'}_i
&=&\sigma(z+\delta+b_i-a_{i'}) \prod_{j(\ne i')}
\frac{\sigma(\delta'+b_i-a_{j})}{\sigma(a_{i'}-a_j)}\\
&\equiv&(-1)^{n-1}\sigma(z+\delta+b_i-a_{i'})(^a_b)^{i'}_i
\end{eqnarray*}
and
$$ [^a_b]^{i'}_i=(^a_b)^{i'}_i\prod_{j(\ne i')}\sigma(a_j-a_{i'}).$$
The corresponding $Y^{i'j'}_{i\;j}$ is,
\begin{eqnarray}
Y^{i'j'}_{i\;j}&=&[^a_b]^{i'}_i[^{a+\hat i'}_{b+\hat i}]^{j'}_j\nonumber\\
&=&\prod_{l(\ne i')}\sigma(\delta'+b_i-a_l)\prod_{m(\ne j')}
  \sigma(\delta'+b'_j-a'_m).
\end{eqnarray}
By using the addition formula Eq.(\ref{add}), we can check that
the solution satisfies Eqs.(\ref{coff-B1})-(\ref{coff-B3})
directly.

  This solution can be
proved to be equivalent with the results obtained by using the
symmetry fusion method for the $A^{(1)}_{n-1}$ model in the
Ref.\cite{jim3}. And it is also equivalent with the evaluation
modules ($n=2$) obtained by Felder and Varchenko in the
Ref.\cite{fe2}.

 Eq.(\ref{solution}) is the only known solution for the form B
coefficients. We do not know if there are  other analytic solutions.
This is still a worthy studying open question.

\section*{Appendix A $\quad$ The relation $F(z)$ between adjacent lattice points}
\setcounter{equation}{0}
\renewcommand{\theequation}{A.\arabic{equation}}
~~ Suppose we go from $(a,b)$ to $(a+\hat i',b+\hat i)$, then we
have $a'_j=a_j+w(\delta_{i'j}-{1\over n}),$
$b'_j=b_j+w(\delta_{ij}-{1\over n})$. From Eq.(\ref{C'-C}) and
(\ref{D'-D}), we may choose
\begin{eqnarray}
C'-C&=&-k_0w(1-\frac{2}{n})+Kw(1-\frac{1}{n}),\label{B.1}\\
D'-D&=&k^0w(1-\frac{2}{n})-Kw(1-\frac{1}{n}) \label{B.2}
\end{eqnarray}
without loss of generality. This is the explicit relations of $C\
D\ (\delta\ E)$ between adjacent lattice points for each form of
$L$-matrices. From Eq.(\ref{L/L-1})
$$ \frac{L(^{a+\hat i'}_{b+\hat i}|z)^{j'}_i}
        {L(^{a}_{b}|z)^{i'}_i}\sim
\frac{\sigma(z+\delta_{i}(i'j')+b_{i}-a_{j'}+w)}
     {\sigma(z+\delta_{i}(i'j')+b_{i}-a_{i'})}$$
and Eq.(\ref{C})
$$ \delta_{i}(i'j')\cong
C-Kb_i+k_0(a_{i'}+a_{j'}),$$ we have
\begin{equation}
\frac{L(^{a+\hat i'}_{b+\hat i}|z)^{j'}_i}
     {L(^{a}_{b}|z)^{i'}_i}\sim
\frac{\sigma(z+C+(1-K)b_i+k_0a_{i'}+(k_0-1)a_{j'}+w)}
     {\sigma(z+C+(1-K)b_i+(k_0-1)a_{i'}+k_0a_{j'})}. \label{B.3}
\end{equation}
The relations of  $F(z)$ and $F'(z)$ (the new function at lattice
point $(a',b')$) can be obtained by putting the explicit forms of
five forms of $L$-matrices (Eqs.(\ref{A1})-(\ref{B})) into
Eq.(\ref{B.3}). For example, we
study the A(1) form.\\
(1) $A(1)\stackrel{i\ i'}{\longrightarrow}A(1) \quad K=1,\ k_0=k^0=0$

From Eq.(\ref{B.1}) and Eq.(\ref{B.2}), one has
\begin{equation} C'= C+w(1-\frac{1}{n}),\quad
D'= D-w(1-\frac{1}{n}).
\end{equation}
 Then Eq.(\ref{A1}) and Eq.(\ref{B.3}) yield
\begin{eqnarray}
&&\frac{L(^{a'}_{b'}|z)^{j'}_i}{L(^{a}_{b}|z)^{i'}_i}\sim
 \frac{\sigma(z+C'-a'_{j'})\sigma(z+D'+b'_{i})F'(z)}
      {\sigma(z+C-a_{i'})\sigma(z+D+b_{i})F(z)}\nonumber\\
&&\quad\quad\quad\quad \sim
 \frac{\sigma(z+C-a_{j'}+w)\sigma(z+D+b_{i})F'(z)}
      {\sigma(z+C-a_{i'})\sigma(z+D+b_{i})F(z)} \nonumber\\
&&\quad\quad\quad\quad \sim
\frac{\sigma(z+C-a_{j'}+w)}{\sigma(z+C-a_{i'})}\nonumber\\
&&\Rightarrow\frac{F'(z)}{F(z)}\sim 1.
\end{eqnarray}
Other A(i)'s are similar. We list them in the following.

 (2) $A(2)\stackrel{i\ i'}{\longrightarrow}A(2) \quad
K=1,\ k_0=0,\
   k^0=1$

 Eq.(\ref{B.1}) and Eq.(\ref{B.2}) give
\begin{equation}
C'= C+w(1-\frac{1}{n}),\quad D'= D-\frac{w}{n}.
\end{equation}
From Eq.(\ref{A2}) and Eq.(\ref{B.3}), we have
\begin{eqnarray*}
\frac{F'(z)}{F(z)}\sim
\frac{\sigma(z+D-b_{i}-w)}{\sigma(z+D-b_{i})}.
\end{eqnarray*}\\
(3) $A(3)\stackrel{i\ i'}{\longrightarrow}A(3) \quad K=1,\
k_0=k^0=1$
\begin{equation}
{F'(z)\over F(z)}\sim {\sigma(z+C+a_{i'}+w)\sigma(z+D-b_i-w)\over
                       \sigma(z+C+a_{i'})\sigma(z+D-b_i)}.
\end{equation}\\
(4) $A(4)\stackrel{i\ i'}{\longrightarrow}A(4) \quad K=1,\ k_0=1,\
k^0=0$
\begin{equation}
{F'(z)\over F(z)}\sim {\sigma(z+C+a_{i'}+w)\over
                       \sigma(z+C+a_{i'})}.
\end{equation}\\
(5) $B\stackrel{i\ i'}{\longrightarrow}B \quad K=0,\ k_0=k^0=0$

From Eq.(\ref{B.1}) and noting $\delta\cong C\cong D$ for this class, we have
$C'\cong D'\cong\delta'\cong C\cong D\cong\delta$. We may choose
$\delta'=\delta$ without loss of generality. Eq.(\ref{B}) and Eq.(\ref{B.3}) imply
\begin{eqnarray}
&&\frac{L(^{a'}_{b'}|z)^{j'}_i}{L(^{a}_{b}|z)^{i'}_i}\sim
\frac{\sigma(z+\delta'+b'_i-a'_{j'})F'(z)}
     {\sigma(z+\delta+b_i-a_{i'})F(z)}\nonumber\\
&&\quad\quad\quad\quad =
\frac{\sigma(z+\delta+b_i-a_{j'}+w)F'(z)}
     {\sigma(z+\delta+b_i-a_{i'})F(z)}\nonumber\\
&&\quad\quad\quad\quad \sim
\frac{\sigma(z+\delta+b_i-a_{j'}+w)}
     {\sigma(z+\delta+b_i-a_{i'})}\nonumber\\
&&\Rightarrow \frac{F'(z)}{F(z)}\sim 1.
\end{eqnarray}

\section*{Appendix B\quad The proof of the theorem 1}
\setcounter{equation}{0}
\renewcommand{\theequation}{B.\arabic{equation}}
~~ To prove the theorem, firstly, we prove the following lemma.

{\bf Lemma 1:} {\sl To any successive product of operators,  if we
transform it by using Eq.(\ref{coff-A}a) and Eq.(\ref{coff-A}b)
such that at each step its inverse order number is reduced (the
adjacent up-indices is exchanged when the left one is bigger than
that of the right one) the final result of the (ab) normal order
expansion is unique.}

Here we assume that in this transformation, two adjacent
operators with same  up-indices do not change the order. And we
think that in every step of the transformation, the location of two
exchanged operators in all terms of the linear combination after
previous step are same.

{\bf Proof:} We can do the procedure by different paths. For
example, if we want to obtain (ab) normal order expansion of
${A^4_{\cdot}}{A^4_{\cdot}}{A^6_{\cdot}}{A^5_{\cdot}}{A^2_{\cdot}}{A^2_{\cdot}}$,
we
may do this in following different paths:\\
(1).  $A^4_\cdot A^4_\cdot A^6_\cdot A^5_\cdot A^2_\cdot
A^2_\cdot\equiv (446522)\stackrel{Q_{4,5}}{\longrightarrow}
(446252)\stackrel{Q_{3,4}}{\longrightarrow}
(442652)\stackrel{Q_{5,6}}{\longrightarrow}
(442625)\stackrel{Q_{4,5}}{\longrightarrow}
(442265)\stackrel{Q_{2,3}}{\longrightarrow}
(424265)\stackrel{Q_{3,4}}{\longrightarrow}
(422465)\stackrel{Q_{1,2}}{\longrightarrow}
(242465)\stackrel{Q_{2,3}}{\longrightarrow}
(224465)\stackrel{Q_{5,6}}{\longrightarrow}
(224456),$ \\
(2). $A^4_\cdot A^4_\cdot A^6_\cdot A^5_\cdot A^2_\cdot
A^2_\cdot\equiv (446522)\stackrel{Q_{3,4}}{\longrightarrow}
(445622)\stackrel{Q_{4,5}}{\longrightarrow}
(445262)\stackrel{Q_{3,4}}{\longrightarrow}
(442562)\stackrel{Q_{2,3}}{\longrightarrow}
(424562)\stackrel{Q_{1,2}}{\longrightarrow}
(244562)\stackrel{Q_{5,6}}{\longrightarrow}
(244526)\stackrel{Q_{4,5}}{\longrightarrow}
(244256)\stackrel{Q_{3,4}}{\longrightarrow}
(242456)\stackrel{Q_{2,3}}{\longrightarrow} (224456)$,\\
where $Q_{i,i+1}$ denotes the exchange of the $i$th operator $A$
and $i+1$th operator $A$ by using rules Eq.(\ref{coff-A}a) and
Eq.(\ref{coff-A}b). We may denote such procedure by product of a
set of exchange operators $\{Q_{i,i+1}\}$ acting on the bunch. For
the path (1) in the example, we have
$$ Q_{5,6}Q_{2,3}Q_{1,2}Q_{3,4}Q_{2,3}Q_{4,5}Q_{5,6}Q_{3,4}Q_{4,5}
A^4_\cdot A^4_\cdot A^6_\cdot A^5_\cdot A^2_\cdot A^2_\cdot
=\sum\cdots A^2_\cdot A^2_\cdot A^4_\cdot A^4_\cdot A^5_\cdot
A^6_\cdot. $$ For the path (2), we have
$$ Q_{2,3}Q_{3,4}Q_{4,5}Q_{5,6}Q_{1,2}Q_{2,3}Q_{3,4}Q_{4,5}Q_{3,4}
A^4_\cdot A^4_\cdot A^6_\cdot A^5_\cdot A^2_\cdot A^2_\cdot
=\sum\cdots A^2_\cdot A^2_\cdot A^4_\cdot A^4_\cdot A^5_\cdot
A^6_\cdot. $$

In general cases, a path of such procedure is denoted by
\begin{eqnarray}
Q_{i_1,i_1+1}Q_{i_2,i_1+1}\cdots Q_{i_s,i_s+1}
\left(A^{j_1}_{k_1}A^{j_2}_{k_2}\cdots A^{j_l}_{k_l}\right)
 =\sum_{j'k'}c^{t_1\cdots
k'_1\cdots}_{j_1\cdots
k_1\cdots}A^{j_{t_1}}_{k'_1}A^{j_{t_2}}_{k'_2}\cdots
A^{j_{t_l}}_{k'_l}         \label{Q-c}
\end{eqnarray}
with $j_{t_1}\leq j_{t_2}\leq\cdots\leq j_{t_l}$. Note that the
original arrangement $\{j_1j_2\cdots j_l\}$ and the final
arrangement $\{j_{t_1}j_{t_2}\cdots j_{t_l}\}$ are same for
whatever path of the (ab) normal product expansion we choose.

Assume there is another path for (ab) normal product expansion
\begin{eqnarray}
Q_{i'_1,i'_1+1}Q_{i'_2,i'_1+1}\cdots Q_{i'_s,i'_s+1}
\left(A^{j_1}_{k_1}A^{j_2}_{k_2}\cdots
A^{j_l}_{k_l}\right)
 =\sum_{j'k'}d^{t_1\cdots
k'_1\cdots}_{j_1\cdots
k_1\cdots}A^{j_{t_1}}_{k'_1}A^{j_{t_2}}_{k'_2}\cdots
A^{j_{t_l}}_{k'_l}.           \label{Q-d}
\end{eqnarray}
Consider the corresponding two products of exchange operators in
the permutation group
$$ P^{(1)}=P_{i_1,i_1+1}P_{i_2,i_2+1}\cdots P_{i_s,i_s+1}$$
and
$$ P^{(2)}=P_{i'_1,i'_1+1}P_{i'_2,i'_2+1}\cdots P_{i'_s,i'_s+1}.$$
They must all be able to permute the arrangement $\{j_1\cdots
j_l\}$ into $\{j_{t_1}j_{t_2}\cdots j_{t_l}\}$. Although some of
the $j$'s may be the same, the permutation $\{^{1\ 2\cdots
l}_{t_1t_2\cdots t_l}\}$ is unique however. This is due to the
rule we do not exchange adjacent operators with same upper
indices. In permutation group, we can express an arbitrary element
by product of exchange operators in different ways. However, we
can always make them equal step by step using the following
equations.
\begin{eqnarray}
&& P_{i,i+1}P_{i,i+1}=id,\label{P-1}\\
&& P_{i,i+1}P_{j,j+1}=P_{j,j+1}P_{i,i+1}\quad (i+1<j),\label{P-2}\\
&&
P_{i,i+1}P_{i+1,i+2}P_{i,i+1}=P_{i+1,i+2}P_{i,i+1}P_{i+1,i+2}.\label{P-3}
\end{eqnarray}
Thus $P^{(1)}$ can be changed to $P^{(2)}$  by using these
equations step by step.

On the other hand, the $\{Q_{i,i+1}\}$ operators have the same
properties. We have checked
\begin{equation}
Q_{i,i+1}Q_{i,i+1}=id  \label{Q-1}
\end{equation}
for two adjacent operators $A^{j_1}_{k_1}A^{j_2}_{k_2'}\ (j_1\neq
j_2)$, and thus it is also valid for all bunches due to
distribution law. We also have
\begin{equation}
Q_{i,i+1}Q_{j,j+1}=Q_{j,j+1}Q_{i,i+1}\quad (i+1<j) \label{Q-2}
\end{equation}
because of the distribution law. Finally we have
\begin{equation}
Q_{i,i+1}Q_{i+1,i+2}Q_{i,i+1}=Q_{i+1,i+2}Q_{i,i+1}Q_{i+1,i+2}
              \label{Q-3}
\end{equation}
due to YBE for any polynomial
$A^{j_1}_{k_1}A^{j_2}_{k_2}A^{j_3}_{k_3}$ with different indices.
Due to distribution law, this equation is also true for any bunch.
Therefore, we can also change
$Q^{(1)}=Q_{i_1,i_1+1}Q_{i_2,i_2+1}\cdots Q_{i_s,i_s+1}$ into
$Q^{(2)}=Q_{i'_1,i'_1+1}Q_{i'_2,i'_2+1}\cdots Q_{i'_s,i'_s+1}$ in
Eq.(\ref{Q-c}) and Eq.(\ref{Q-d}), respectively, by using
Eqs.(\ref{Q-1})-(\ref{Q-3}) step by step since $P^{(1)}$ and
$P^{(2)}$ can be equaled in such way by using
Eqs.(\ref{P-1})-(\ref{P-3}), respectively. Thus we have
$c^{j_{t_1}\cdots k'_1\cdots}_{j_{1}\cdots k_1\cdots}
=d^{j_{t_1}\cdots k'_1\cdots}_{j_{1}\cdots k_1\cdots}.$

 We then conclude that the resulting (ab) normal order expansion
of the two paths give the same result. Therefore, all paths give
the same result. $\quad\quad {\bf \Delta}$

Corollaries then follows:

{\bf Corollary 1:} {\sl If in a product of successive product of
operators $CA^{i'}_iA^{j'}_jD$ where $C,D$ are all products of
operators, we obtain the combination of $CA^{j'}_\cdot
A^{i'}_\cdot D$ (it is, $C(\alpha A^{j'}_iA^{i'}_j+\beta
A^{j'}_jA^{i'}_i)D)$ by changing ( with rule (ab) in
Eq.(\ref{coff-A})) two adjacent operators whose  up-indices are
unequal, the results of their (ab) normal order expansions are
same, if the procedure is done according to the rules described in
lemma 1.}

{\bf Proof:} If $i'>j'$, we can regard this changing procedure as the
first step of the (ab) normal order expansion. Thus, we can
prove it. If $i'<j'$, we can do the (ab) normal order expansion of
 $C(\alpha A^{j'}_iA^{i'}_j+\beta A^{j'}_jA^{i'}_i)D$,
 and let the first step as the changing of
$A^{j'}_\cdot A^{i'}_\cdot$ into $A^{i'}_\cdot A^{j'}_\cdot$.
Then, By using the rule (ab), we can prove that $i'j'\rightarrow
j'i'\rightarrow i'j'$ is the identical transformation. So with the
distributive law, the (ab) normal order expansion of bunch
$C(\alpha A^{j'}_iA^{i'}_j+\beta A^{j'}_jA^{i'}_i)D$ =the (ab)
normal order expansion of $CA^{i'}_iA^{j'}_jD$. Therefore, this
corollary is proved.\quad\quad ${\bf \Delta}$

{\bf Corollary 2:} {\sl With the rules of the Eq.(\ref{coff-A}a)
and Eq.(\ref{coff-A}b), if a polynomial (a linear combination of
products) of operators $C$ can be changed to $D$
$(C\stackrel{(ab)}{\rightarrow} D)$, the (ab) normal order
expansions of $C$ and $D$ are same, if the expansion is done
according to the rules described in lemma 1.}

{\bf Proof:} Because each step of the transformation do not affect
the result of the expansion. \quad\quad${\bf \Delta}$

 Thus the Eq.(\ref{coff-A}a) Eq.(\ref{coff-A}b) are compatible with the
(ab) normal order expansion and the (abc)  normal order expansion.

Here we note that same results of the (ab) normal order expansion give
same
results of the (abc) normal order expansion, so the above two
corollaries are also true for the (abc) normal order expansion.

Next, we prove the following lemma.

{\bf Lemma 2:} {\sl The (abc) normal order expansion of the bunch
$CA^{i'}_jA^{i'}_kD$ and the bunch $CA^{i'}_kA^{i'}_jD$ are same.}

{\bf Proof:} We need only to prove this when they are monomials.
We prove the following propositions by using the mathematical
inductive
method:\\
Proposition (i). This lemma is true when the inverse order number
is
zero.\\
Proposition (ii). If the lemma is true when the inverse order number is
smaller than $m$, it is also true when the inverse order number is equal
to $m$.

 The first proposition is obvious, because in this case,
$CA^{i'}_jA^{i'}_kD$ and $CA^{i'}_kA^{i'}_jD$ are all (ab) normal
order products. To obtain the (abc) normalization, we only need to
rearrange the down-indices of the part of the product where the
up-indices are same from the smaller to the bigger by rule
Eq.(\ref{coff-A}c). Both of the bunches have same sets of the
down-indices for up-indices $i'$. Therefore, the (abc) normal
order products of them  are same.

To the second proposition, we have the following cases:

($\alpha$). If in $C$ or $D$, we can rearrange the up-indices $\{i'\}$
of them to reduce the inverse order number, for example,
$D\stackrel{(ab)}{\longrightarrow}D'$. We can obtain $CA^{i'}_jA^{i'}_kD'$
and $CA^{i'}_kA^{i'}_jD'$. According to the corollary 2 of the lemma 1,
the (ab) normal order expansions of both of them will keep unchanged.
However, because the inverse order number must be smaller that $m$ now, so
according to assumption of the proposition (ii), their (abc) normal order
expansions are same. Therefore, the (abc) normal order expansions of the
$CA^{i'}_jA^{i'}_kD$ and the $CA^{i'}_kA^{i'}_jD$ are same.

($\beta$). If $C$ and $D$ have already been normalized but the
inverse order number of the bunch as a whole can be reduced,
namely, the bunch is not an (ab) normal order product. We can let
$C=C_1A^{i'_c}_{i_c}, D=A^{i'_d}_{i_d}D_1$. Then we must have
$i'_c>i'$ or (and) $i'>i'_d$. Let us assume $i'_c>i'$. These two
bunches can be rewritten as $T_1=C_1
A^{i'_c}_{i_c}A^{i'}_jA^{i'}_kD$ and $T_2=C_1
A^{i'_c}_{i_c}A^{i'}_kA^{i'}_jD$ respectively. According to the
rule (ab) in Eq.(\ref{coff-A}), we can change them as
$T_1\Rightarrow T'_1=C_1\sum_{rst}a_{rst} A^{i'}_r A^{i'}_s
A^{i'_c}_t D$ and $T_2\Rightarrow T'_2=C_1\sum_{rst}b_{rst}
A^{i'}_r A^{i'}_s A^{i'_c}_t D$, where $a_{rst}$ and $b_{rst}$ are
some coefficients.  With the help of the YBE which we studied in
section 5, one can see that these two combinations
$\sum_{rst}a_{rst} A^{i'}_r A^{i'}_s A^{i'_c}_t$  and
$\sum_{rst}b_{rst} A^{i'}_r A^{i'}_s A^{i'_c}_t$ are the same if
we take the rule Eq.(\ref{coff-A}c) into account. Thus we must
have
\begin{eqnarray*}
\sum_{rst}a_{rst}A^{i'}_r A^{i'}_s A^{i'_c}_t -
\sum_{rst}b_{rst}A^{i'}_r A^{i'}_s A^{i'_c}_t
=\sum_t\left(\sum_{rs}(a_{rst}-b_{rst})A^{i'}_r
A^{i'}_s\right)A^{i'_c}_t
\end{eqnarray*}
with $\sum_{rs}(a_{rst}-b_{rst})A^{i'}_r A^{i'}_s=0$ if we take
the rule Eq.(\ref{coff-A}c) into account. This is to say
\begin{equation}
a_{rst}+a_{srt}=b_{rst}+b_{srt}=2c_{rst}\quad \mbox{for each }t.
                  \label{c-rst}
\end{equation}
Thus we have
\begin{eqnarray*}
T'_1=\sum_{rst}C_1a_{rst}A^{i'}_r A^{i'}_s A^{i'_c}_t D \equiv
\sum_t\sum_{rs}\left(C_1a_{rst}A^{i'}_r A^{i'}_s D_t\right)
\end{eqnarray*}
and
\begin{eqnarray*}
T'_2=\sum_{rst}C_1b_{rst}A^{i'}_r A^{i'}_s A^{i'_c}_t D \equiv
\sum_t\sum_{rs}\left(C_1b_{rst}A^{i'}_r A^{i'}_s D_t\right).
\end{eqnarray*}
From Eq.(\ref{c-rst}) and due to the assumption of the proposition
(ii), the (abc) normal order expansions of $T'_1$ and $T'_2$ are
the same. According to the procedure of the (abc) normal order
expansion, we see that the (abc) normal order expansions of $T_1$
and $T_2$ are same.

 If $i'>i_d'$, the proof is similar. So we see that the proposition
(ii) is true.

 Thus, with the mathematical inductive method, we prove the lemma
 2. $\quad\quad {\bf \Delta}$

From the corollary 2 of lemma 1 and lemma 2, we obtain theorem 1.

\end{document}